\documentclass[prb,aps,preprint,showpacs,superscriptaddress]{revtex4}
\usepackage{graphicx}

\begin{document}

\title{A hard metallic material: Osmium Diboride}

\author{Z. Y. Chen}
\affiliation{Hefei National Laboratory for Physical Sciences at
  Microscale,
  University of Science and Technology of
  China, Hefei, Anhui 230026, People's Republic of China}

\affiliation{USTC Shanghai Institute for Advanced Studies,
  University of Science and Technology of China,
  Shanghai 201315, People's Republic of China}

\author{H. J. Xiang}
\affiliation{Hefei National Laboratory for Physical Sciences at
  Microscale,
  University of Science and Technology of
  China, Hefei, Anhui 230026, People's Republic of China}

\affiliation{USTC Shanghai Institute for Advanced Studies,
  University of Science and Technology of China,
  Shanghai 201315, People's Republic of China}

\author{Jinlong Yang}

\thanks{Corresponding author. E-mail: jlyang@ustc.edu.cn}

\affiliation{Hefei National Laboratory for Physical Sciences at
  Microscale,
  University of Science and Technology of
  China, Hefei, Anhui 230026, People's Republic of China}

\affiliation{USTC Shanghai Institute for Advanced Studies,
  University of Science and Technology of China,
  Shanghai 201315, People's Republic of China}

\author{J. G. Hou}
\affiliation{Hefei National Laboratory for Physical Sciences at
  Microscale,
  University of Science and Technology of
  China, Hefei, Anhui 230026, People's Republic of China}

\author{Qingshi Zhu}
\affiliation{Hefei National Laboratory for Physical Sciences at
  Microscale,
  University of Science and Technology of
  China, Hefei, Anhui 230026, People's Republic of China}

\affiliation{USTC Shanghai Institute for Advanced Studies,
  University of Science and Technology of China,
  Shanghai 201315, People's Republic of China}

\date{\today}

\begin{abstract}
 We calculate the structural and electronic properties of OsB$_{2}$
 using density functional theory with or without taking into account
 spin-orbit (SO) interaction. Our results show that the bulk modulus
 with and without SO interaction are 364 and 365 Gpa respectively,
 both are in good agreement with experiment (365-395 Gpa).
The evidence of covalent bonding of Os-B, which plays an important
role to form a hard material, is indicated both in charge
density, atoms in molecules analysis, and density of states
analysis. The good metallicity and
hardness of OsB$_{2}$ might suggest its potential application as
hard conductors.
\end{abstract}

\pacs{71.15.Mb, 71.20.-b, 62.20.Dc, 81.05.Bx}

\maketitle

Hard materials are of great interest due to their
superior properties of higher compressional strength, thermal
conductivity, refractive index, and chemical stability beside
higher hardness.\cite{appl} Besides the two well-known super-hard
materials of diamond and cubic boron nitride ($c$-BN),
experimental and theoretical efforts are devoted to searching a
new class of hard meterials. Covalent materials are much better
candidates for high hardness than ionic compounds because
electrostatic interactions are omnidirectional and yield low
bond-bending force constants, which result in low shear
modulus.\cite{cov, sci} Some potential hard materials such as C3N4
polymorphs,\cite{c3n4} Si3N4,\cite{si3n4} BC2N,\cite{bc2n} and
B6O,\cite{b6o} have been proposed and investigated intensively in
recent years.

The pure Osmium (Os) has an exceptionally high bulk modulus
(395-462 GPa) while its hardness is of only 400
kg/mm$^{2}$.\cite{os} The high bulk modulus of Os mainly due to
the high valence electron density, while the low hardness is
related to the metallic bonds and the hexagonal close-packed (HCP)
crystal structure of Os. It has been found that many transition
metals, such as W, Ru, and Zr, which are soft in their pure
metals, can be converted into hard materials by combining with
small covalent bond-forming atoms such as boron, carbon, oxygen,
or nitrogen.\cite{wc1, wc2, RuO, wc3}
For Os related materials, OsO$_{2}$ has been predicted to be very
stiff,\cite{RuO} and some compounds
like OsB$_{0.11}$, OsB$_{2}$, and Os$_{2}$B$_{3}$ have already
been synthesized.\cite{osb,osb2geom}
Recently, Cumberland {\it et al.} studied experimentally the
mechanical properties of the OsB$_{2}$ compound, and showed that
it is an ultra-incompressible (365-395 GPa) and hard
($\geq$2000kg/mm$^{2}$) material.\cite{exp,sci}

In this brief report, we carry out density functional calculations
on the electronic properties and the bulk modulus of OsB$_{2}$,
using the highly accurate full potential linearized augmented
plane wave plus local orbital (APW + LO) method as implemented in
the Wien2k code.\cite{wien2k} The nonoverlapping muffin-tin (MT)
sphere radiis of 2.02 and 1.57 bohr are used for the Os and B
atoms, respectively. Expansion in spherical harmonics for the
radial wave functions are taken up to l=10. The value of
R$_{MT}$K$_{max}$ (the smallest muffin tin radius multiplied by
the maximum k value in the expansion of plane waves in the basis
set) which determines the accuracy of the basis set used is set to
$7.0$. The total Brillouin zones are sampled with 6000 k points,
i.e., 672 k points in the irreducible wedge of the Brillouin zone.
For the exchange-correlation energy functional, we use the local
density approximation (LDA) in our calculations. Since Os is a
heavy element, a fully relativistic calculation is performed for
core states, whereas the valence states are treated in both scalar
and fully relativistic fashions. The spin-orbit coupling (SOC) is
self-consistently added via the second variational step scheme
with relativistic p$_{1/2}$ local orbitals included.\cite{kune} We
find there is no spontaneous spin polarization in OsB$_{2}$ in the
framework of the local spin density approximation (LSDA) or
LSDA+SO methods. Hereafter for the sake of simplicity, we only
discussed the results from the spin-restricted calculations.

\begin{figure}[!hbp]
  \caption{(Color online) The crystal structure for OsB$_{2}$. The lattice vectors are
    denoted as $\bf{a}$, $\bf{b}$, and $\bf{c}$.}
    \label{fig1}
  \includegraphics[width=6.0cm]{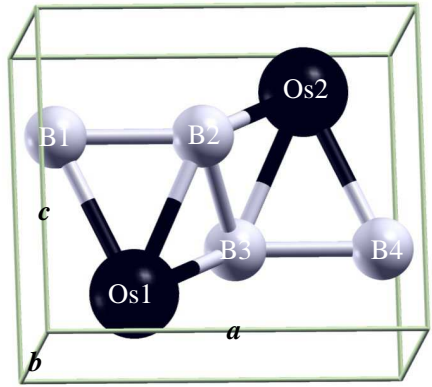}
\end{figure}

OsB$_2$ has an orthorhombic lattice (space group $Pmmn$, No. 59)
with the experimental lattice parameters ${\bf a}=4.6832 \pm
0.0001 $, ${\bf b}=2.8717 \pm 0.0002 $, and ${\bf c}=4.0761 \pm
0.0001 $ \AA. In the orthorhombic structure, two Os atoms occupy
the 2a Wyckoff sites and four B atoms occupy the 4f sites, as
shown in Fig.~\ref{fig1}. We optimize both lattice and internal
coordinates to get a relaxed structure for OsB$_2$.
The total energy per unit cell as a
function of volume is calculated and the results are fitted using the
Murnaghan equation of state to get the bulk
modulus.\cite{murna}
The corresponding equilibrium lattice parameters, bulk modulus
from both LDA and LDA+SO methods are compared with the
experimental result, as shown in Table~\ref{table1}. We can see
that the equilibrium lattice parameters calculated both in the
absence and in the presence of SO are very close with the
experimental value. And the bulk modulus calculated in the absence
and in the presence of SO are 365 and 364 Gpa, respectively, both
are in good agreement with the experimental result (365-395
GPa).\cite{exp}

\begin{figure}[!hbp]
  \caption{(Color online) (a) The valence charge density contour (in
    e/a.u.$^{3}$) plot in the plane including Os1, Os2, B2, and B3.
    (b) and (c) show respectively the valence charge density and
    difference charge density  contour (in e/a.u.$^{3}$) plot in the
    plane including Os1, B1, and B2.
    The charge density shown here is the result from
    the LDA+SO calculation. The LDA charge density has similar character.
    Os1, Os2, B1, B2, and B3 refer to the atoms labeled in Fig.~\ref{fig1}(a).}
  \label{fig2}
  \includegraphics[width=6cm]{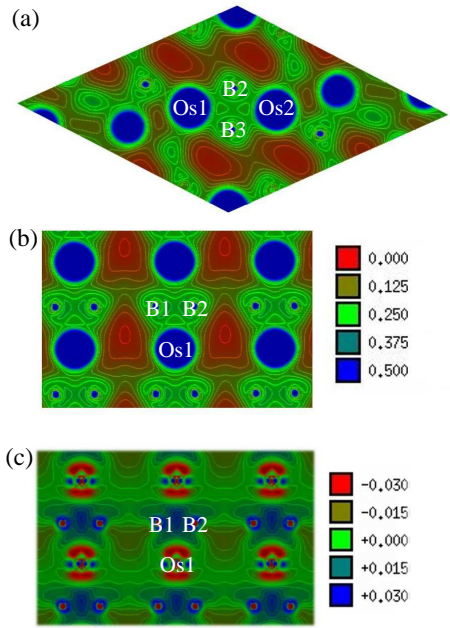}
\end{figure}

As discussed above, the hardness of a material is largely
determined by the bonding type in the system. Thus the electronic
properties of OsB$_2$ such as the electron density, band
structure, and the density of state (DOS) are calculated in the
absence and in the presence of SO at the equilibrium geometries.
We find that the electron density from the LDA calculation is
almost the same as that from the LDA+SO calculation. Here we show
the electron density for OsB$_2$ from the LDA+SO calculation in
Fig.~\ref{fig2}. Two charge density contour plots (one in the
plane including Os1, Os2, B2, and B3, and the other in the plane
including Os1, B1, and B2) are shown in Fig.~\ref{fig2}(a) and
(b), respectively. There are some electrons between Os atoms and
theirs neighbor B atoms, indicating covalent bonding in OsB$_2$.
To gain a more detailed insight into the bonding characters of
OsB$_2$, we plot the difference density (the difference between
total density and superposition of atomic densities) in
Fig.~\ref{fig2}(c). We can see that some electrons are transferred
from both B and Os atoms to B-B bonds and Os-B bonds. Clearly, two
neighbor B atoms form very strong covalent bond. And the bonding
between the Os atom and its neighbor B atoms is also covalent. The
fact that Os-B bond is covalent rather than ionic is also evident
from Bader's atoms in molecules (AIM) theory,\cite{aim} which
indicates there is only about 0.05 electron transferred from an Os
atom to B atoms. Moreover, there is some region where the charge
density is rather low, implying the bonding in this system is
directional but not metallic, as can be seen from
Fig.~\ref{fig2}(a) and (b). Since there are eight B atoms around
an Os atom, forming a three dimensional network instead of a
plane, the anisotropy in OsB$_2$ is not expected to be very large.
As discussed above, the strong covalent bonding of Os-B plays an
important role to form a hard material, since the highly
directional bonding is needed to withstand both elastic and
plastic deformations. OsB$_{2}$ might be potential superhard
material for effectively cutting ferrous metals, including steel,
since diamond reacts with steel producing iron carbide.\cite{sci}
A recent report showed that the hardness of the Fe-based alloys is
less than 1300kg/mm$^{2}$,\cite{Fe} which is far less than that of
the OsB$_{2}$($\geq$2000kg/mm$^{2}$).\cite{exp}

\begin{figure}[!hbp]
  \caption{(a) Band structure of OsB$_{2}$ from the LDA
  calculation. (b) Band structure of OsB$_{2}$ from the LDA+SO
  calculation. $\Gamma=(0,0,0)$,
  $M=(0,0.5,0)$, $N=(0.5,0.5,0)$, $K=(0.5,0,0)$, $A=(0,0,0.5)$.}
  \label{fig3}
  \includegraphics[width=6cm]{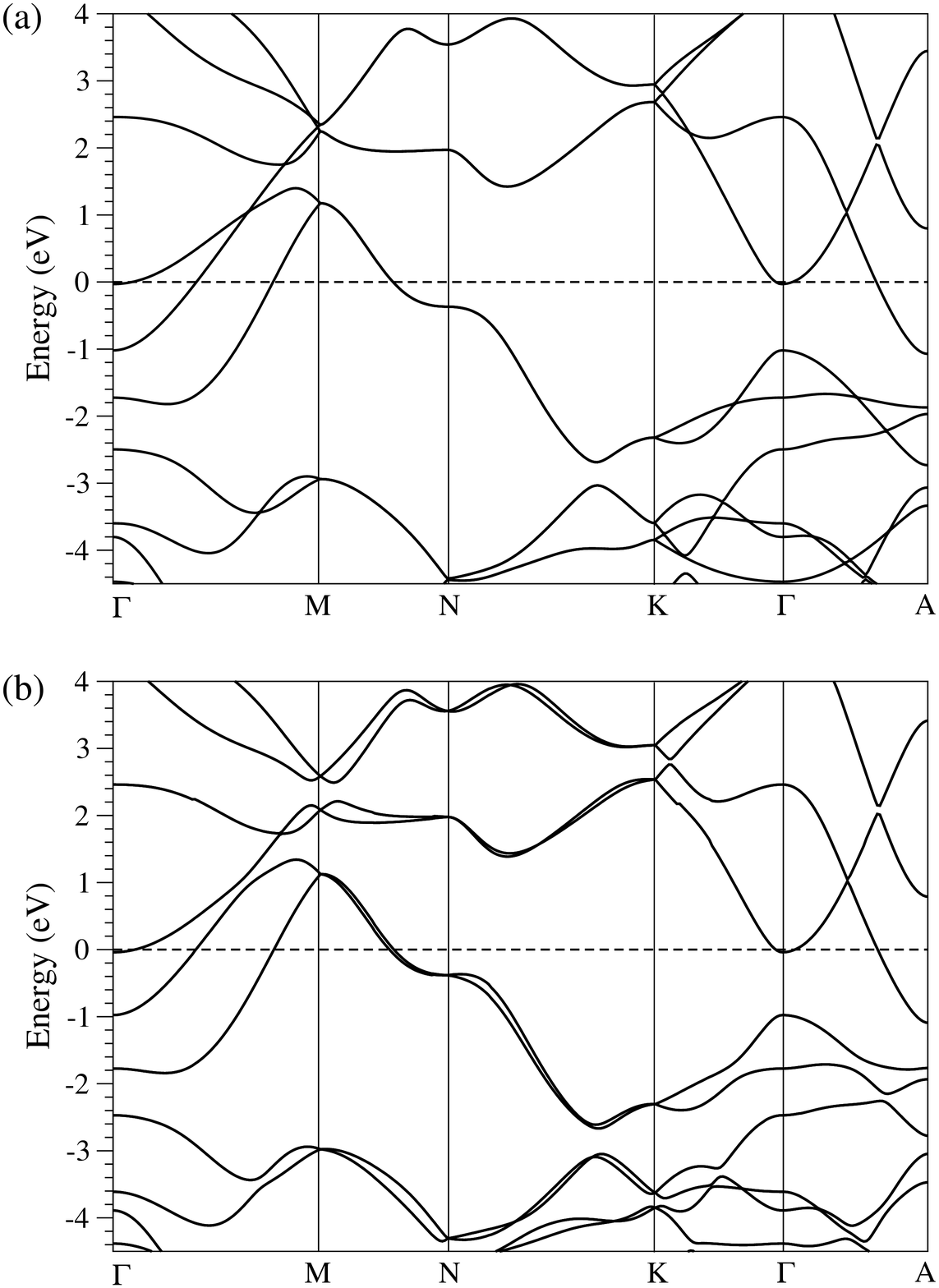}
\end{figure}

\begin{figure}[!hbp]
  \caption{(Color online) Partial and total DOS of OsB$_{2}$ from the
  LDA and LDA+SO calculations.}
  \label{fig4}
  \includegraphics[width=7cm]{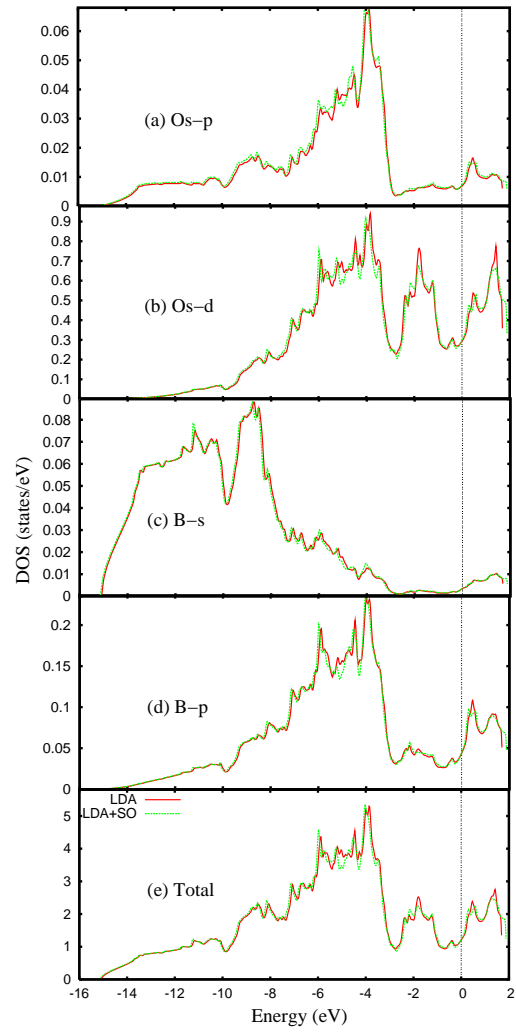}
\end{figure}

Although the LDA electron density is almost indiscernible from
that from the LDA+SO calculation, we note that there are some
differences in the band structures and DOS between two results.
Fig.~\ref{fig3}(a) and (b) show the band structures near the Fermi
energy in the absence or presence of SO, respectively. Due to the
spin-orbit interaction, some bands are split: some bands crossing
each other in Fig.~\ref{fig3}(a) no longer cross in
Fig.~\ref{fig3}(b), and more noticeably almost all bands in the
direction of M-N-K are split in the LDA+SO result. Except the band
splitting in the presence of SO, the both band structures have
main features in common. Both Fig.~\ref{fig3}(a) and (b) show that
the bands are dispersive but not flat indicating not large
anisotropy in OsB$_{2}$. We note that there are several bands
crossing the Fermi level which are rather dispersive in the whole
Brillouin zone, indicating good electronic mobility in OsB$_{2}$.
The metallicity is uncommon in superhard materials. Most of the
superhard materials are insulators or semicondunctor.\cite{wc3}
This special character of OsB$_{2}$ should bring it special
application on electron conductivity. Fig.~\ref{fig4} show the
total and partial DOS with or without SOC which are obtained using
a modified tetrahedron method of Bl\"ochl {\it et al.}.\cite{tetr}
We can see some minor differences between the results with or
without SOC. Fig.~\ref{fig4}(e) also implies the good metallicity
due to the substaintially large total DOS at the Fermi level. From
the calculated partial DOS in Fig.~\ref{fig4}, one can see that
the electronic structure of OsB$_{2}$ is governed by a strong
hybridization between the Os-d and B-p states, while with a rather
small contribution from the Os-p and B-s states. This strong
hybridization of Os-d and B-p also indicates the strong covalent
bonding of Os-B.

Besides the bulk modulus, we want to mention that the shear modulus is also
relevant for the non-cubic OsB$_{2}$ materials.
The shear modulus can be computed from the elastic constants.
Unfortunately, the force calculation when including the spin-orbit coupling has
not been implemented in the Wien2k package yet. As an
approximation, in the calculation of elastic constants, we keep
the relative positions of the atoms fixed at small elastic strains of
the lattice. Both calculations without SOC or with SOC are carried
out. The procedure for computing elastic constants is similar to that
used by Zhou {\it et al.}.\cite{OsB2_PAW}
For comparison, we also compute the elastic constants using VASP\cite{vasp} with
LDA but no SOC included.
In these calculations, we perform two series of calculations with
atomic internal coordinates fixed or relaxed, respectively.
Our results reported in Table~\ref{table2}.
We can see that the Voigt shear modulus $G$  calculated
using the Wien2k package (235.1
and 251.8 GPa when including SOC or not, respectively.) are
in consistent with that (253.6 GPa) calculated using VASP with fixed
atomic internal coordinates, but are much larger than that (197.1 GPa)
calculated using VASP with a relaxation of the atomic internal coordinates.
The results produced by VASP are in good agreement with those reported
by Zhou {\it et al.}.\cite{OsB2_PAW}
Thus we find that is not a very good approximate that
computing the shear modulus with atomic internal coordinates fixed.
The accurate shear modulus with SOC included might be computed in our future
works.

In conclusion, the mechanical and electronic properties of
OsB$_{2}$ are studied using the highly accurate full potential
linearized APW + LO method. The OsB$_{2}$ is found to be metallic,
which is uncommon in hard materials. The calculated bulk modulus
with and without SOC are 364 and 365 GPa respectively, which are
in good agreement with the experiment (365-395GPa).\cite{exp} The
evidence of covalent bonding of Os-B, which plays an important
role to form a hard material, is manifested both in the difference
charge density, AIM analysis, and DOS analysis. Our results
indicate that OsB$_{2}$ might serve as hard conductors.

This work is partially supported by the National Project for the
Development of Key Fundamental Sciences in China (G1999075305,
G2001CB3095), by the National Natural Science Foundation of China
(50121202, 10474087, 20533030), by the USTC-HP HPC project, and by
the Virtual Laboratory for Computational Chemistry of CNIC and
Supercomputing Center of CNIC, Chinese Academy of Sciences.

\newpage

\begin{table}
  \caption{Structural parameters and bulk modulus of OsB$_{2}$.}
  \label{table1}
  \begin{tabular}{cccc}
    \hline\hline
    & LDA & LDA+SO & Experiment \\
    \hline
    Lattice parameters(\AA):& & &\\
    $\bf a$&4.6433&4.6581&4.6832$\pm$0.0001 \cite{osb2geom}\\
    $\bf b$&2.8467&2.8700&2.8717$\pm$0.0002 \cite{osb2geom}\\
    $\bf c$&4.0432&4.0560&4.0761$\pm$0.0001 \cite{osb2geom}\\
    Bulk modulus(GPa)&365&364&365-395 \cite{exp}\\
    \hline\hline
  \end{tabular}
\end{table}

\begin{table}
  \caption{Calculated elastic constants and Voigt shear modulus $G$ of
    orthorhombic OsB$_{2}$. Unless otherwise stated, all results are
    computed with atomic internal coordinates unrelaxed.
    All values are in units of GPa. }
  \label{table2}
  \begin{tabular}{cccccc}
    \hline\hline
    & Wien2K LDA & Wien2K LDA+SO & VASP LDA & VASP LDA (relaxed) & Other VASP
    LDA (relaxed)\cite{OsB2_PAW} \\
    \hline
    $C_{11}$   & 628.9& 608.5& 622.4& 597.2& 597.0 \\
    $C_{12}$   & 194.7& 198.3& 185.8& 188.7& 198.1 \\
    $C_{13}$   & 235.0& 220.9& 225.3& 217.5& 206.1 \\
    $C_{22}$   & 627.8& 590.3& 624.9& 584.5& 581.2 \\
    $C_{23}$   & 126.8& 129.9& 124.1& 164.0& 142.6 \\
    $C_{33}$   & 923.2& 855.7& 910.4& 833.8& 825.0 \\
    $C_{44}$   & 185.5& 175.1& 182.0&  80.2& 70.1  \\
    $C_{55}$   & 313.5& 292.6& 311.4& 214.5& 212.0 \\
    $C_{66}$   & 218.6& 205.9& 233.9& 209.2& 201.3 \\
    $G$ & 251.8        &235.1       & 253.6        &197.1        &193.8    \\
    \hline\hline
  \end{tabular}
\end{table}

\end{document}